\documentclass{ifacconf}

\makeatletter
	\let\cl@part\relax
	\let\old@ssect\@ssect 
\makeatother

\usepackage[utf8]{inputenc}
\usepackage{amsmath}
\usepackage{array}

\usepackage{amsfonts}
\usepackage[draft]{minted}
\usepackage{graphicx}
\usepackage{mathtools}
\usepackage{nicefrac}
\usepackage{booktabs}
\usepackage{multirow}
\usepackage{csquotes}
\usepackage[usenames,dvipsnames]{xcolor}
\usepackage[
	colorlinks = true,
	linkcolor = Fuchsia,
	urlcolor  = JungleGreen,
	citecolor = magenta
]{hyperref}
\usepackage[capitalise,noabbrev]{cleveref} 

\usepackage{color}

\usepackage[all]{nowidow}
\usepackage{hyphenat}
\usepackage{enumitem}
\usepackage{siunitx}

\usepackage{natbib}

\graphicspath{{./graphics/}}

\ifdefined\draft%
	\usepackage{epstopdf}
	\DeclareGraphicsExtensions{.png,.pdf}
\fi

\makeatletter
	\def\@ssect#1#2#3#4#5#6{%
		\NR@gettitle{#6}
		\old@ssect{#1}{#2}{#3}{#4}{#5}{#6}
	}
\makeatother

\begin{document}

	\begin{frontmatter}

		\title{Analysis of a Dynamic Voluntary Contribution Mechanism Public Good Game\thanksref{footnoteinfo}}

\thanks[footnoteinfo]{This is an exteneded version of the paper that has been published in IPE Journal, Volume 26, 2017 \citep{bogatov-ipe-journal-2017}.}

\author{Dmytro Bogatov}

\address{Worcester Polytechnic Institute, Worcester, MA 01609 USA \\ (e-mail: dbogatov@wpi.edu)}

		\begin{abstract}
			I present a dynamic, voluntary contribution mechanism, public good game and derive its potential outcomes.
In each period, players endogenously determine contribution productivity by engaging in costly investment.
The level of contribution productivity carries from period to period, creating a dynamic link between periods.
The investment mimics investing in the stock of technology for producing public goods such as national defense or a clean environment.
After investing, players decide how much of their remaining money to contribute to provision of the public good, as in traditional public good games.
I analyze three kinds of outcomes of the game: the lowest payoff outcome, the Nash Equilibria, and socially optimal behavior.
In the lowest payoff outcome, all players receive payoffs of zero.
Nash Equilibrium occurs when players invest any amount and contribute all or nothing depending on the contribution productivity.
Therefore, there are infinitely many Nash Equilibria strategies.
Finally, the socially optimal result occurs when players invest everything in early periods, then at some point switch to contributing everything.
My goal is to discover and explain this point.
I use mathematical analysis and computer simulation to derive the results.

		\end{abstract}


	\end{frontmatter}

	\section{Introduction}

	In this paper, I present a dynamic, voluntary contribution mechanism, public good game and solve it for the lowest payoff, Nash Equilibria and socially optimal outcomes.
	In the lowest payoff scenario, a team gets the smallest possible payoff.
	In equilibrium case, team players act in their own interest in the light of what everyone else is doing.
	While the lowest payoff and Nash Equilibria cases are relatively straightforward, I concentrate on the socially optimal outcome.
	In that case, players act in a team's interest to maximize team's payoff.
	To derive the solution for the socially optimal outcome, I build a mathematical model, simplify it, then use computational methods and regression analysis to derive a generic analytical solution.

	Public goods are goods that are non-excludable and non-rival in consumption. Non-excludability means that people cannot be excluded or restricted from using the good.
	Non-rivalry means that the use of the good by one individual does not reduce its availability to others.
	One simple example of a public good is a hurricane siren in a small town.
	No one can be excluded from using the siren and the use of the siren does not affect its availability to other people.
	Another example of a public good is an open sourced, open licensed software.
	It is almost impossible to exclude people from using the software and the use of the software by anyone does not affect its availability to anyone else.

	The challenge associated with the voluntary provision of public goods is the free rider problem.
	The problem occurs when those who benefit from a public good or service do not pay for it or do not contribute to it.
	This leads to the non-production or under-provision of a good or service.
	As long as this problem is not solved, the voluntary provision of the public good may be ineffective and unsuccessful.
	Consider the example of national defense.
	This good is public since it is neither rival nor excludable.
	The free rider problem may occur if a part of the society decides not to pay for it.
	The good will remain public, but provision of the good will be hurt due to the lack of contributions.
	Moreover, the other people, those who pay, will have an incentive to stop paying too as they might not want to pay for ``free riders''.

	Consider an example of a public good that needs contributions in order to operate.
	Wikipedia is an open platform where people can post articles and other people can read them.
	While the service is free, it does need resources, technology and infrastructure to operate.
	Wikipedia needs thousands of highly qualified encyclopedists, millions of dollars and a team of software engineers to function.
	All of these resources are provided voluntarily, thus, keeping the good non-excludable.

	It is important to solve or mitigate the problem of ``free riders'' as it may have negative consequences.
	As more people start to ``free ride'', the incentive to pay for the good decreases and even more people prefer not to contribute.
	This self-reinforcing process will lead to overconsumption, exhaustion or even destruction of the public good.
	At certain point, the system or service will not have enough resources to operate.

	It is important to understand people's behavior when they decide whether or not to contribute to a public good.
	It is always tempting to avoid contributing, thus, become a free rider.
	To understand this behavior it is helpful to analyze how the equilibrium may be achieved, what the lowest possible payoff is, and how to achieve socially optimal behavior.
	This analysis is a primary goal of the paper.

	A number of experiments were conducted on the free rider problem and public good provision.
	\citeauthor{free-ride} test a strong free rider hypothesis by a series of 12 experiments and find it not applicable to real world situations \citep{free-ride}.
	They tested two versions of free rider hypothesis --- weak and strong --- with experiments.
	``[T]he `weak' version of the free-rider hypothesis, \ldots{} states that the voluntary provision of public goods by groups will be sub-optimal and the `strong' version, \ldots{} argues that (virtually) no public goods at all will be provided through voluntary means.''
	They have found that strong version of free rider hypothesis is not practically supported, while the `weak' version is \citep{free-ride}.

	\citeauthor{free-ride-divergent} attempt to reconcile divergent experimental results of other papers --- those which concluded that individuals always free ride a lot, free ride a little or never free ride.
	They conclude that contribution decision depends on the conditions of the experiment \citep{free-ride-divergent}.
	\citeauthor{ledyard-chpater} has a chapter in his textbook dedicated to public good games experiments.
	He carefully describes how such experiments need to be conducted, and how certain aspects of experimental setup affect the end results \citep{ledyard-chpater}.
	Finally, \citeauthor{sustaining-cooperation} surveys a large number of papers on public good games in his work.
	He analyses the effectiveness of monetary and non-monetary punishments on sustaining cooperation in public good games.
	He concludes that monetary punishments are generally effective, however, there are certain caveats such as cost-effectiveness of the punishment itself and the issue of ``anti-social'' punishments.
	He concludes that there are other, non-monetary, punitive measures such as expressions of disapproval or social exclusion \citep{sustaining-cooperation}.

	To analyze people's decision-making regarding contributing to public goods, I have examined the game of \citep{ngo-paper}.
	In the following sections I will describe the game and its rules, and provide an example of one period (round) in the game.
	I will consider three possible outcomes --- the lowest payoff, Nash Equilibria strategies and socially optimal behavior, concentrating on the latter one.
	While analyzing the socially optimal behavior, I will construct a mathematical model first, solve it computationally and apply regression analysis tools to derive a generic solution.
	I will then validate the solution mathematically.
	At the end, I will summarize the paper.

	\section{The Game}

	The game \citep{ngo-paper} occurs among groups of $4$ people and consists of $10$ periods.
	Each period has two stages: an investment stage and a contribution stage.
	At the start of every period, all players receive endowments of $10$.

	\subsection{Investment Stage}

		This game incorporates endogenous determination of contribution productivity and dynamic links between periods.
		These two aspects are incorporated through the investment stage of each period where players have the opportunity to increase their contribution productivity from the starting value of $0.30$ (the low used by \citep{group-size-effects}).
		First, players vote to determine the amount each player in the group will invest in increasing contribution productivity by choosing a whole number between zero and ten, inclusive.
		A median voter rule is applied and the group's investment is the average of the two middle votes.
		Then, contribution productivity increases by $0.01$ multiplied by the investment.
		For example, in the first period where contribution productivity equals $0.30$, if players vote $1$, $3$, $5$, and $6$, the investment will be $4$.
		Therefore, all four group members invest $4$ and have a remaining amount of $6$ left after the investment.
		Contribution productivity then equals $0.34$.
		Players vote on an investment and invest in contribution productivity every period, and the amount builds throughout the ten periods.

		\[
			\text{Contribution productivity: } M_t = M_{t-1} + 0.01 \cdot I_t
		\]
		\[
			\text{for } t \in [1, 10]
		\]
		\[
			M_0 = 0.3
		\]

	\subsection{Contribution Stage}

		Following the investment stage is the contribution stage where players decide how to allocate their remaining money between private consumption and public good, similar to a standard public good game.
		If the investment had been $10$, then the players must contribute zero to the public good because they have no money left in the period.
		The sum of the group members' contributions is multiplied by the new $M_t$ each period and this amount, in addition to any money the player has remaining after the investment and contribution stage, is the player's payoff for that period (note that every player benefits from the sum of the group contributions multiplied by the $M_t$ regardless of whether or not they contribute).
		Thus, each player's payoff for each period is:

		\[
			\pi_{it} = \omega - I_t - c_{it} + M_t \sum c_{jt}
		\]

		where:
			\begin{itemize}
				\item
					$\pi_{i}$ is the individual's payoff for the period,
				\item
					$\omega$ is the individual's endowment,
				\item
					$I$ is the investment for the period,
				\item
					$c_i$ is the individual's contribution to the public good,
				\item
					$M$ is the contribution productivity,
				\item
					$c_j$ is the other members' contributions to the public good, and
				\item
					$t$ is the period.
			\end{itemize}

		See an example in \cref{table:example} ($M_0 = 0.3$).

		\begin{table}[!ht]
	\caption{Example game}%
	\label{table:example}
	\begin{tabular*}{\linewidth}{ !{\extracolsep\fill} c c c c c c c } 
		\toprule
		Players	& $\omega$	& $I_t$	& $M_t$		& $C_{it}$	& $M_t \sum c_{jt}$	& $\pi_{it}$	\\
		\midrule
		1		& $10$		& $3$	& $0.33$	& $7$		& $4.95$			& $4.95$		\\
		2		& $10$		& $3$	& $0.33$	& $5$		& $4.95$			& $6.95$		\\
		3		& $10$		& $3$	& $0.33$	& $3$		& $4.95$			& $8.95$		\\
		4		& $10$		& $3$	& $0.33$	& $0$		& $4.95$			& $11.95$		\\
		\bottomrule
	\end{tabular*}
\end{table}

		In this case, the players' endowments are $10$.
		They collectively decide to invest $3$.
		Their multiplier gets increased by $0.01 \cdot 3 = 0.03$ and is equal to $0.33$.
		Players individually decide to contribute $7$ (all that is left), $5$, $3$ and nothing.
		The payoffs are $4.95$, $6.95$, $8.95$ and $11.95$.

		The player $4$ is a clear example of a free rider.
		This player did not contribute anything relying on other's contributions, yet still got a benefit from public good.
		Although this behavior is legitimate in the game, it is clearly not socially optimal.

	\section{Potential Outcomes}

	In this section, I will analyze three potential outcomes of the game.
	The most straightforward scenario is the one where a team gets the lowest possible payoff.
	The next part is a set of equilibrium scenarios in which each player acts in his own interest in light of what everyone else is doing.
	In this case, a player has some beliefs about other players' strategies and acts accordingly to maximize his payoff.
	Finally, I will analyze a scenario where players maximize the team's payoff.
	This is a socially optimal behavior where players act in the team's interests, not their own.
	In each of these scenarios, I will identify the team's strategy and the payoff a team can get.

	\subsection{Lowest payoff outcome}

		The lowest payoff one could get is $0$.
		This occurs if the group invests everything in every period and never contributes anything.
		Payoffs are $0$ in every period and $0$ at the end of $10$ periods.

\begin{figure}
	\centering
	\includegraphics[width=\linewidth]{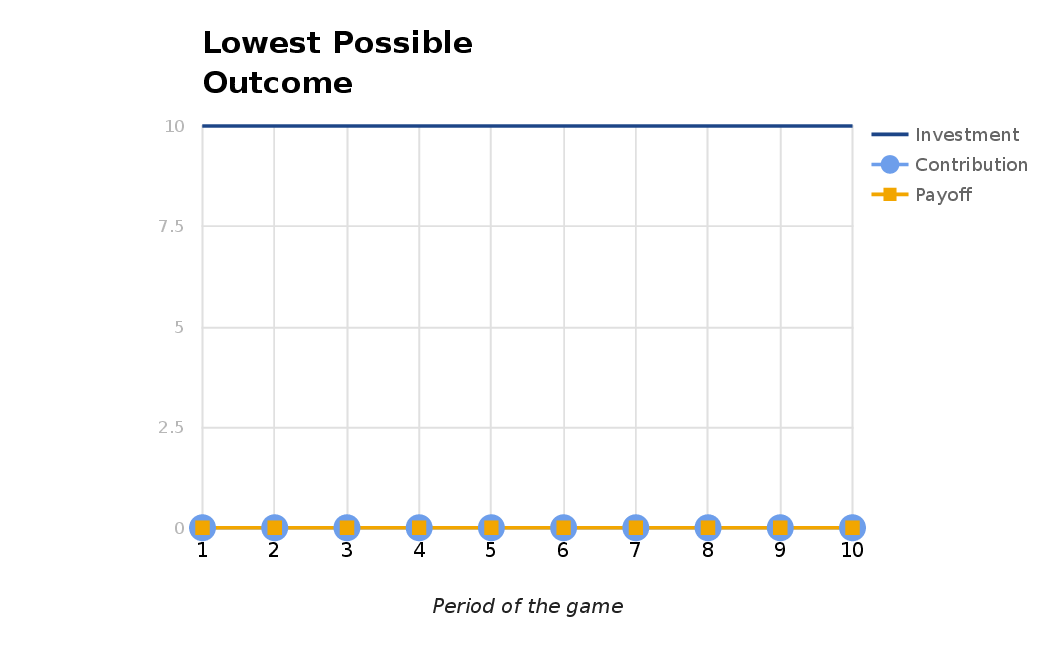}
	\caption{Lowest possible payoff}%
	\label{figure:lowest}
\end{figure}

		The reason this outcome is poor is because the investment is never productively used.
		Investments are meaningless without contributions, because the purpose of investments is to amplify contributions to achieve higher returns.
		See \cref{figure:lowest}.

	\subsection{Nash equilibrium}

		The formal definition of Nash Equilibrium as defined by John Nash is the following:

		\begin{displayquote}
			\emph{
				Equilibrium point is $n$-tuple strategy such that each player's mixed strategy maximizes his payoff if the strategies of the others are held fixed.
				Thus, each player's strategy is optimal against those of the others.
			}
			\citep{non-cooperative-games}
		\end{displayquote}

		A less formal definition is given by \citeauthor{intro-to-applicable-game-theory}:

		\begin{displayquote}
			\emph{
				The collection of strategies in which each player's predicted strategy must be that player's best response to the predicted strategies of the other players.
			}
			\citep{intro-to-applicable-game-theory}
		\end{displayquote}

		My goal for the set of equilibrium scenarios is to find each player's strategy such that the player acts in his own interests in light of what other players are doing.
		Since the game is symmetric (each player starts the game in the same conditions), the players will share the same strategy.

		\subsubsection{General strategy}

			According to the definition of Nash equilibrium, each player will aim to maximize his payoff.
			During the contribution stage the optimal actions are:

			\[
				c_{it}
					\begin{cases}
						= 0 					& \text{if} \; M_t < 1 \\
						\in [0, \omega - I_t]	& \text{if} \; M_t = 1 \\
						= \omega - I_t			& \text{if} \; M_t > 1
					\end{cases}
			\]

			To paraphrase, in equilibrium, each player's individual strategy during the contribution stage is to contribute nothing if the multiplier is below $1$, contribute any amount if the multiplier is equal to $1$, or contribute everything the player has if the multiplier is higher than $1$.
			Please note that when $M = 1$, any contribution will lead to the same payoff.

			These actions are optimal because when $M < 1$ a player's payoff will be less than the value he contributed, if no one else contributed.
			The player will be worse off if he contributes while others do not (or ``free ride'').
			Therefore, acting in his own interest, he does not contribute.
			When $M \ge 1$ a player's payoff will be larger or equal to the value he contributed, so he contributes all he has regardless of other players' actions.

			During the investment stage, however, any decision will satisfy equilibrium strategy since regardless of the voting, all players make the same investment.
			For example, if players vote $1$, $3$, $5$ and $7$, a median voter rule will produce $4$ and \emph{all players} will invest $4$.
			Since all players make the same investment, no player can be in a beneficial position relative to other players at the end of the investment stage.
			Therefore, any investment decision will lead to an equilibrium.
			Since a player may make any investment, there are an infinite\footnote{
				If we restrict the smallest possible amount of investment --- for example to $\$0.01$ --- then there will be a finite number of solutions.
			} number of equilibrium solutions.

		\subsubsection{Special cases}

			Although there are an infinite number of solutions to the equilibrium scenarios, I identified three special cases.
			Notice that in equilibrium scenarios all players get the same payoff.

\begin{figure}
	\begin{center}
		\includegraphics[width=\linewidth]{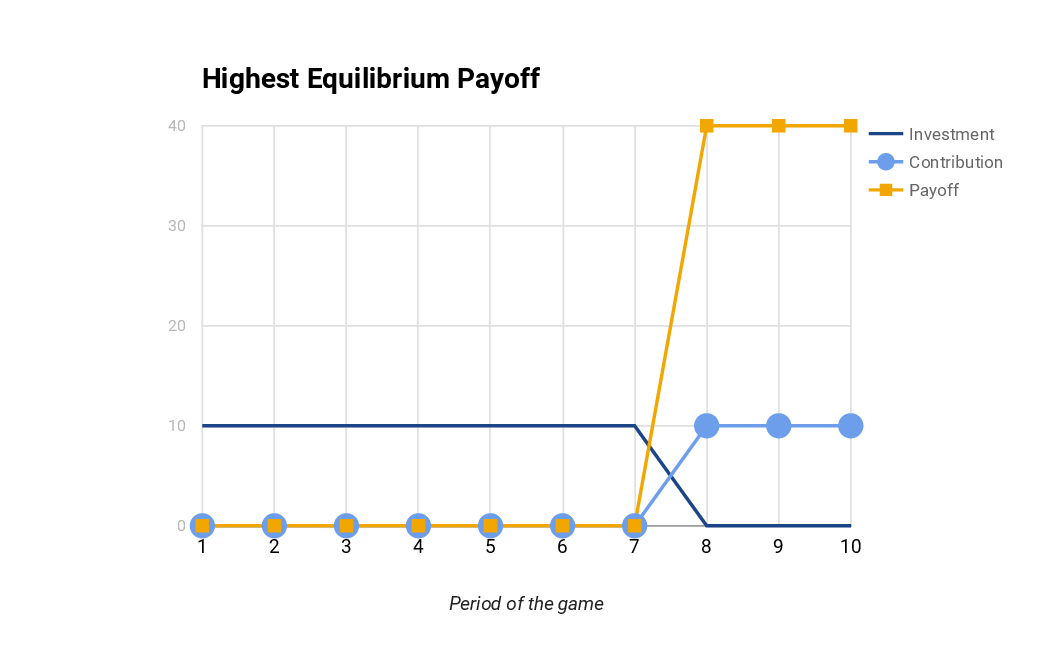}
		\caption{Highest possible equilibrium payoff}%
		\label{figure:nash-highest}
	\end{center}
\end{figure}

			The highest payoff a player can get is $120$.
			This occurs if players invest everything until the $8^\text{th}$ period when $M = 1$.
			Starting from the period $8$ and until the end of the game, players contribute their full endowments $\omega = 10$.
			Each period their payoff is $1 \cdot 4 \cdot 10 = 40$ and at the end they get $3 \cdot 40 = 120$.
			See \cref{figure:nash-highest}.

\begin{figure}
	\begin{center}
		\includegraphics[width=\linewidth]{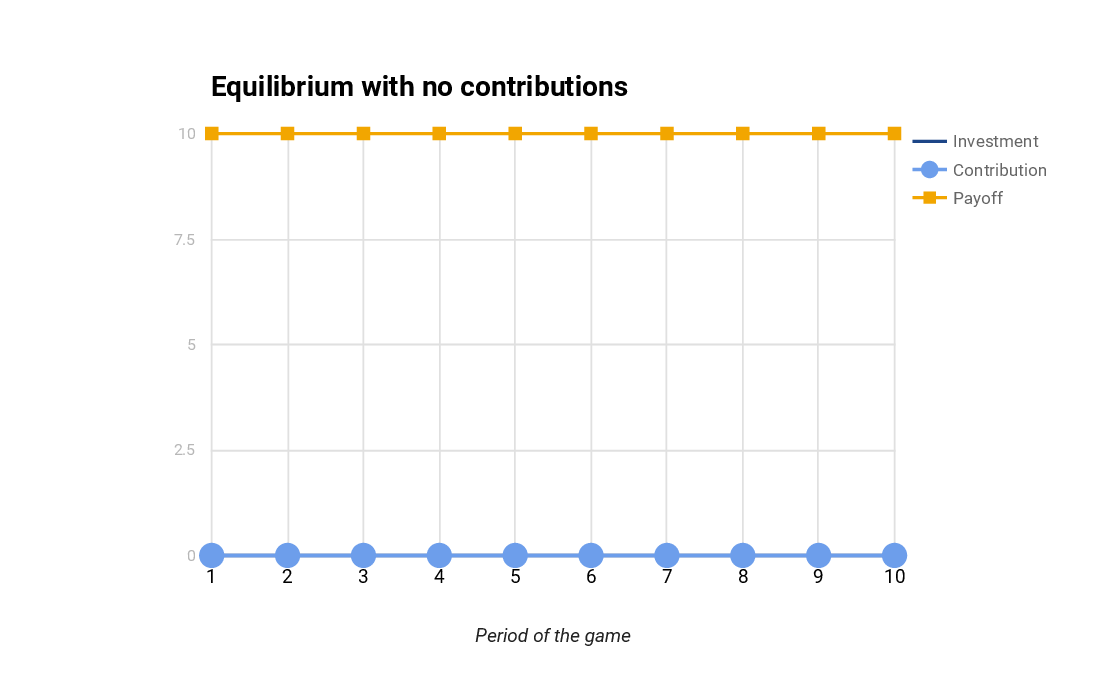}
		\caption{Equilibrium payoff with no investments and no contributions}%
		\label{figure:nash-nocontrib}
	\end{center}
\end{figure}

			If players decide to invest nothing in all periods, they will keep their endowments and get $10 \cdot 10 = 100$.
			See \cref{figure:nash-nocontrib}.

\begin{figure}
	\begin{center}
		\includegraphics[width=\linewidth]{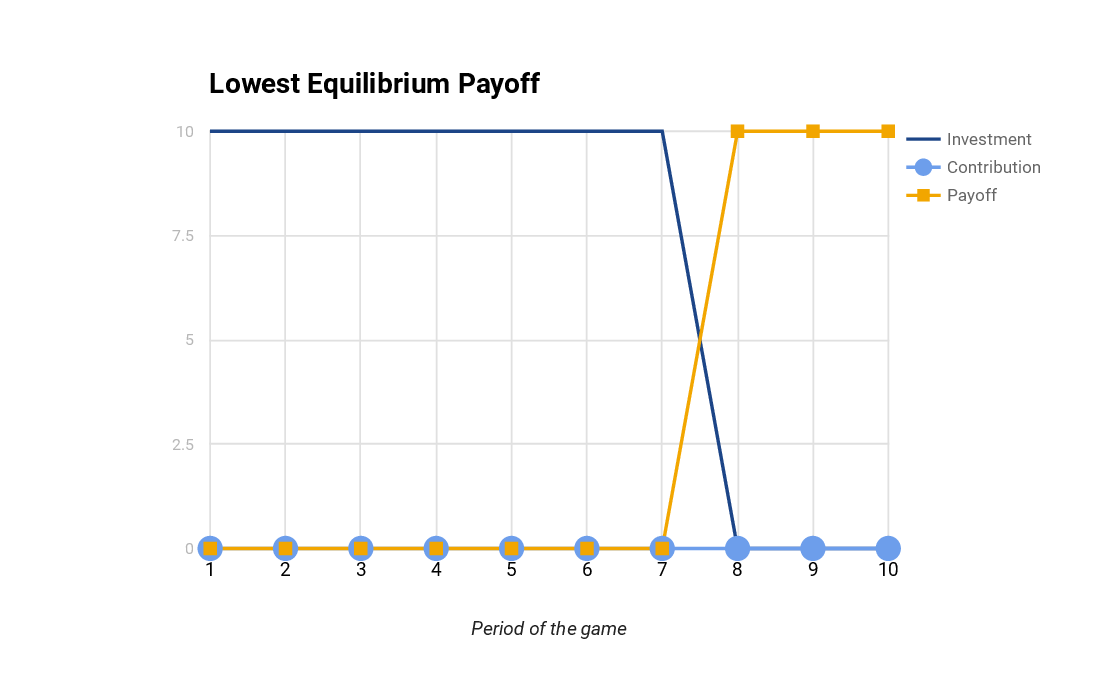}
		\caption{Lowest possible equilibrium payoff}%
		\label{figure:nash-lowest}
	\end{center}
\end{figure}

			The lowest amount a player can get is $30$.
			This occurs if players build the multiplier equal to $1$, and afterwards do not contribute.
			It will take $7$ periods to build such multiplier.
			Players will keep their endowments last $3$ periods and will get $3 \cdot 10 = 30$.
			See \cref{figure:nash-lowest}.

			\[
				\begin{cases}
					I_t = 10 	& \text{for } t \in [1, 7] \\
					c_{it} = 0	& \text{for } \forall i \; \forall t
				\end{cases}
			\]

			Notice that an individual player cannot affect such scenario --- if the other three players vote the same amount to invest, that amount gets invested.

	\subsection{Socially optimal behavior}

		\subsubsection{The mathematical model}

			To approach this problem I have to build a mathematical model.
			Thinking of a single period I can define a function, $f$, of investment, $I$, and contribution, $C$, that returns a payoff.
			Let me define this function for the period after the $t_\text{th}$ period.

			\[
				f(I, C) = [M_t \cdot 4 \cdot C] + [\omega - C - I]
			\]
			\[
				M_t = M_{t-1} \cdot (1 + 0.01 \cdot I)
			\]
			\[
				M_0 = 0.3
			\]

			$M_t \cdot 4 \cdot C$ is payoff and $\omega - C - I$ is the amount left after both stages.

			Since every $\$1$ contributed will give at least $0.3 \cdot 4 = \$1.2$ back, keeping money in the contribution is not socially optimal.
			A player will always be better off contributing a unit than keeping it.
			Therefore, I can state a fact, cosnsitent with a deeper analysis of the game by \citep[Appendix]{ngo-paper}:

			\begin{displayquote}
				\emph{
					The optimal result requires contributing all that is left after the investment.
				}
			\end{displayquote}

			Now, the term $\omega - C - I$ equals $0$, so I see that $I = \omega - C$, which means that I can eliminate one of the two variables --- $C$ or $I$.	Let me instead introduce a new variable, $p$, as a fraction of endowment which player invests.

			Now $I = p \cdot \omega$ and $C = (1-p) \cdot \omega$.
			Let me redefine the function:

			\[
				f(p) = M_t \cdot 4 \cdot \omega \cdot (1 - p)
			\]
			\[
				M_t = M_{t-1} \cdot (1 + 0.01 \cdot \omega \cdot p)
			\]
			\[
				M_0 = 0.3
			\]
			where:
			\begin{itemize}
				\item
					$p$ is the \emph{proportion} of investment,
				\item
					$\omega$ is the endowment ($10$), and
				\item
					$M_t$ is the $t_\text{th}$ multiplier.
			\end{itemize}

			From now, let me solve it specifically for my case, when endowment is $10$.
			\[
				f(p) = 40 \cdot M_t \cdot (1 - p)
			\]
			\[
				M_t = M_{t-1} \cdot (1 + \frac{p}{10})
			\]
			\[
				M_0 = 0.3
			\]
			where:
			\begin{itemize}
				\item
					$p$ is the \emph{proportion} of investment, and
				\item
					$M_t$ is the $t_\text{th}$ multiplier
			\end{itemize}

			This is a recursive function that is difficult to solve analytically.
			One way to solve it anaytically is to reformulate it without recursion, see \cref{section:analytical-solution}.

		\subsubsection{The computational model}

			Having the mathematical model, I can make use of a computer to solve it numerically, and then use regression analysis to derive the analytical solution.

			First, let me approximate how much time it would take for the computer to execute this simulation.

			If I ran the simulation for $10$ periods with a step $p$ that is at least as small as $0.1$, the time complexity of the algorithm would be $\mathcal{O}\left( n^a \right)$, which is $\approx 10^{10} \cdot c$ computations.
			It would take months to run the simulation on a usual PC, so I need to improve the formula.

			Let me then state the second fact, again cosnsitent with a deeper analysis of the game by \citep[Appendix]{ngo-paper}:
			\begin{displayquote}
				\emph{
					The optimal solution requires that players first only invest then only contribute.
					In other words, $p = 1.0$ for some number of periods, then $p \in [0.0, 1.0]$ for one period,
					and finally $p = 0.0$ for the rest of the game.
				}
			\end{displayquote}

			This finding is based on the fact that the value of an investment declines and the value of a contribution grows with time.

			This constrint reduces the complexity of the algorithm to linear --- $\mathcal{O}(n)$.

			Let me define two functions (see \cref{appendix:code}) that compute payoff for each possible $p$.
			This program runs the game $10$ periods for all possible values of $p$ taking into account the second assumption.
			For each such run it returns a payoff.
			The result of the computation is in \cref{appendix:output}.

\begin{figure}
	\begin{center}
		\includegraphics[width=\linewidth]{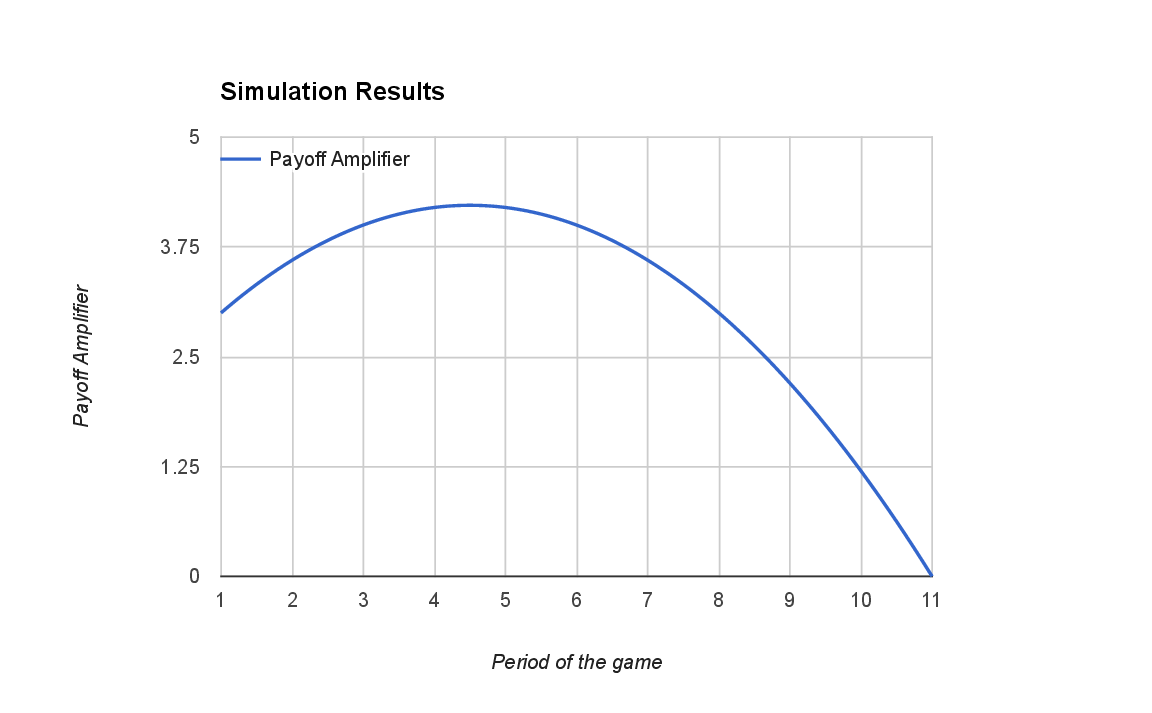}
		\caption{Function of payoff depending on time when a player switches to contribution --- $f(p)$}%
		\label{figure:function}
	\end{center}
\end{figure}

			Although this simulation immediately gives me the value of $p$ where payoff is maximized, I am interested in a generic solution.
			Let me plot the function.
			According to \cref{figure:function}, the resulting function has a global maximum and is probably quadratic with $a < 0$.

		\subsubsection{Regression Analysis}

			The last step is to try to ``estimate'' the formula for the data.
			I enter the data in the regression tool\footnote{\url{http://www.xuru.org/rt/pr.asp}}, give it a potential mathematical model (in my case it is a quadratic formula $ax^2 + bx + c$) and run it.
			The tool would estimate the formula with numeric coefficients and error value.
			In this case, I got a quadratic formula with the error value equal to $0$, which means perfect fit.
			Now, let me derive a generic formula so that it fits numeric one.
			This is an artistic process --- I play around with numbers trying to compose variables such that they fit the quadratic formula.

			\begin{multline}\label{equation:function}
				f(x) = 400 \cdot \\
				\left[ - m \cdot \omega \cdot x^2 + (m \cdot \omega \cdot T - M_0) \cdot x + M_0 \cdot T \right]
			\end{multline}
			\[
				x_\text{max} = \frac{T}{2} - \frac{M_0}{2 \cdot m \cdot \omega}
			\]
			\[
				f_\text{max} = f(x_\text{max}) = f \left(\frac{T}{2} - \frac{M_0}{2 \cdot m \cdot \omega} \right)
			\]
			where:
			\begin{itemize}
				\item
					$m$ is the increase in multiplier $0.01$,
				\item
					$T$ is the number of periods, and
				\item
					$x$ is the stage when players switch to contributing.
					The number before the decimal point defines a period.
					The number after the decimal point defines an investment in that period.
			\end{itemize}

			The computation and regression analysis were conducted for different values of $m$, $\omega$, $M_t$, $T$ and $x$ to demonstrate the robustness of the formula.

			As a final step, with a generic formula I can compute the actual outcome by using my specific initial values.

			\[
				f(x) = -0.1 x^2 + 0.7 x + 3
			\]
			\[
				x_{\text{optimal}} = 3.5
			\]

			which indicates investment until the $4_\text{th}$ period and in that period investment of $0.5$

			\[
				f_{\text{optimal}} = f(x_{\text{optimal}}) = 169
			\]

			which implies the payoff of $169$.

\begin{figure}
	\begin{center}
		\includegraphics[width=\linewidth]{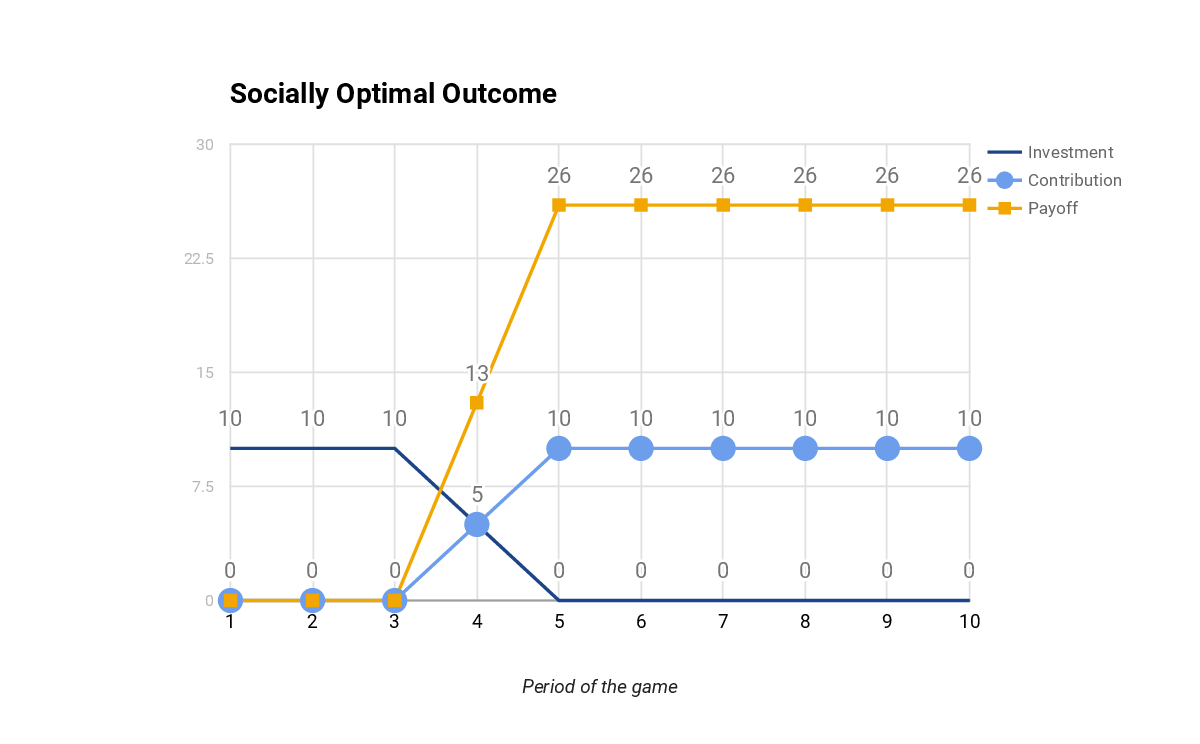}
		\caption{Socially optimal outcome payoff}%
		\label{figure:social}
	\end{center}
\end{figure}

			The maximum payoff for each player in the group of $4$ is $169$.
			To achieve this, the players must invest everything until period $4$,
			then invest half of their endowments, and contribute everything afterwards.
			See \cref{figure:social}.

		\subsection{Analytical Solution}\label{section:analytical-solution}

			This section was inspired by the \citep{ngo-paper} and was added after the publication in IPE 2017.

			If $x$ is the number of \emph{full} rounds of investment (the same definition as in the erlier sections), then the payoff for each round is $400 \cdot (M_0 - m \cdot \omega \cdot x)$.
			Since there are $T - x$ such rounds, the final function becomes:
			\begin{align*}
				f(x)	& = 400 \cdot (M_0 - m \cdot \omega \cdot x)(T - x) \\
						& = 400 \cdot \left[ - m \cdot \omega \cdot x^2 + (m \cdot \omega \cdot T - M_0) \cdot x + M_0 \cdot T \right]
			\end{align*}

			Note, that the analytically derived solution matches the earlier results of regression analysis, see \cref{equation:function}.

	\section{Conclusions}

	In this paper I have analyzed three potential outcomes of the game.
	The lowest possible outcome occurs if players only invest and never contribute.
	A player's payoff is $0$ since it is unreasonable to invest in contribution multiplier and not to contribute.
	The equilibrium strategies occur if all players contribute all or nothing depending on the contribution productivity.
	Finally, I have analyzed the socially optimal outcome.
	It occurs if players invest until a certain period, then only contribute.
	I have analytically derived a generic formula that produces the optimal strategy.

	\section{Acknowledgements}

	I cannot express enough thanks to my colleague Jacqueline Ngo for her contributions and cooperation.
	It is her game that I analyze, so this paper could not have been produced without her work \citep{ngo-paper}.

	I am grateful to my advisor and supervisor Dr.\ Alexander Smith for the time and effort he put into this paper.
	I appreciate his guidance and support.

	I would also like to thank Dr.\ Oleg Pavlov for motivating me to start working on the paper in the first place.
	It was his idea to expand a little extra credit project into a conference paper, which then evolved into a journal paper.

	Last but not least, I would like to thank Worcester Polytechnic Institute Social Sciences Department for funding my trip to the IPE Conference in Washington, DC\@.
	Presenting at the top level conference has given me necessary experience and motivation to complete this paper.

	\bibliography{bibfile}

	\appendix

	\onecolumn

		\section{Computational Solution Code}\label{appendix:code}
\begin{minted}{swift}
import Foundation

func arrayToString(p : [Double]) -> String {
	var result : String = "";

	for e in p {
		result += "\(e), ";
	}

	return result;
}

func compute(p : Double...) -> Double {
	return compute(p);
}

func compute(p : [Double]) -> Double {
	var result : Double = 0.0;
	var m : Double = 0.3;

	for var i = 0; i < p.count; i++ {
		m += p[i] * 0.1;
		result += m * (1 - p[i]);
	}

	return result;
}

func run(rounds : Int, step : Double) -> String {
	var best : Double = 0.0;
	var p = [Double](count: rounds, repeatedValue: 0.0);
	var result : String = "";

	for i in 0...(rounds-1) {

		print("Changing \(i+1)-th proportion");

		for p[i] = 0.0; p[i] < 1.0; p[i] += step {

			let res = compute(p);
			result += "\(Double(i) + p[i])\t\(res)\n";

			if res > best {
				best = res;
			}

			print("\tres: \(String(format: "%.3f", res));\
			\tvalues: \(arrayToString(p))");
		}

		p[i] = 1.0;
	}

	print("Result");
	print("\tBest value: \(best)");
	print("\tMax payout: \(4 * best) USD");

	return result;
}

run(10, step: 0.1);
\end{minted}

		\section{Computational Results}\label{appendix:output}
\begin{minted}{text}
Changing 1-th proportion
	res: 3.000;	values: 0.0, 0.0, 0.0, 0.0, 0.0, 0.0, 0.0, 0.0, 0.0, 0.0,
	res: 3.069;	values: 0.1, 0.0, 0.0, 0.0, 0.0, 0.0, 0.0, 0.0, 0.0, 0.0,
	res: 3.136;	values: 0.2, 0.0, 0.0, 0.0, 0.0, 0.0, 0.0, 0.0, 0.0, 0.0,
	res: 3.201;	values: 0.3, 0.0, 0.0, 0.0, 0.0, 0.0, 0.0, 0.0, 0.0, 0.0,
	res: 3.264;	values: 0.4, 0.0, 0.0, 0.0, 0.0, 0.0, 0.0, 0.0, 0.0, 0.0,
	res: 3.325;	values: 0.5, 0.0, 0.0, 0.0, 0.0, 0.0, 0.0, 0.0, 0.0, 0.0,
	res: 3.384;	values: 0.6, 0.0, 0.0, 0.0, 0.0, 0.0, 0.0, 0.0, 0.0, 0.0,
	res: 3.441;	values: 0.7, 0.0, 0.0, 0.0, 0.0, 0.0, 0.0, 0.0, 0.0, 0.0,
	res: 3.496;	values: 0.8, 0.0, 0.0, 0.0, 0.0, 0.0, 0.0, 0.0, 0.0, 0.0,
	res: 3.549;	values: 0.9, 0.0, 0.0, 0.0, 0.0, 0.0, 0.0, 0.0, 0.0, 0.0,
	res: 3.600;	values: 1.0, 0.0, 0.0, 0.0, 0.0, 0.0, 0.0, 0.0, 0.0, 0.0,
Changing 2-th proportion
	res: 3.600;	values: 1.0, 0.0, 0.0, 0.0, 0.0, 0.0, 0.0, 0.0, 0.0, 0.0,
	res: 3.649;	values: 1.0, 0.1, 0.0, 0.0, 0.0, 0.0, 0.0, 0.0, 0.0, 0.0,
	res: 3.696;	values: 1.0, 0.2, 0.0, 0.0, 0.0, 0.0, 0.0, 0.0, 0.0, 0.0,
	res: 3.741;	values: 1.0, 0.3, 0.0, 0.0, 0.0, 0.0, 0.0, 0.0, 0.0, 0.0,
	res: 3.784;	values: 1.0, 0.4, 0.0, 0.0, 0.0, 0.0, 0.0, 0.0, 0.0, 0.0,
	res: 3.825;	values: 1.0, 0.5, 0.0, 0.0, 0.0, 0.0, 0.0, 0.0, 0.0, 0.0,
	res: 3.864;	values: 1.0, 0.6, 0.0, 0.0, 0.0, 0.0, 0.0, 0.0, 0.0, 0.0,
	res: 3.901;	values: 1.0, 0.7, 0.0, 0.0, 0.0, 0.0, 0.0, 0.0, 0.0, 0.0,
	res: 3.936;	values: 1.0, 0.8, 0.0, 0.0, 0.0, 0.0, 0.0, 0.0, 0.0, 0.0,
	res: 3.969;	values: 1.0, 0.9, 0.0, 0.0, 0.0, 0.0, 0.0, 0.0, 0.0, 0.0,
	res: 4.000;	values: 1.0, 1.0, 0.0, 0.0, 0.0, 0.0, 0.0, 0.0, 0.0, 0.0,
Changing 3-th proportion
	res: 4.000;	values: 1.0, 1.0, 0.0, 0.0, 0.0, 0.0, 0.0, 0.0, 0.0, 0.0,
	res: 4.029;	values: 1.0, 1.0, 0.1, 0.0, 0.0, 0.0, 0.0, 0.0, 0.0, 0.0,
	res: 4.056;	values: 1.0, 1.0, 0.2, 0.0, 0.0, 0.0, 0.0, 0.0, 0.0, 0.0,
	res: 4.081;	values: 1.0, 1.0, 0.3, 0.0, 0.0, 0.0, 0.0, 0.0, 0.0, 0.0,
	res: 4.104;	values: 1.0, 1.0, 0.4, 0.0, 0.0, 0.0, 0.0, 0.0, 0.0, 0.0,
	res: 4.125;	values: 1.0, 1.0, 0.5, 0.0, 0.0, 0.0, 0.0, 0.0, 0.0, 0.0,
	res: 4.144;	values: 1.0, 1.0, 0.6, 0.0, 0.0, 0.0, 0.0, 0.0, 0.0, 0.0,
	res: 4.161;	values: 1.0, 1.0, 0.7, 0.0, 0.0, 0.0, 0.0, 0.0, 0.0, 0.0,
	res: 4.176;	values: 1.0, 1.0, 0.8, 0.0, 0.0, 0.0, 0.0, 0.0, 0.0, 0.0,
	res: 4.189;	values: 1.0, 1.0, 0.9, 0.0, 0.0, 0.0, 0.0, 0.0, 0.0, 0.0,
	res: 4.200;	values: 1.0, 1.0, 1.0, 0.0, 0.0, 0.0, 0.0, 0.0, 0.0, 0.0,
Changing 4-th proportion
	res: 4.200;	values: 1.0, 1.0, 1.0, 0.0, 0.0, 0.0, 0.0, 0.0, 0.0, 0.0,
	res: 4.209;	values: 1.0, 1.0, 1.0, 0.1, 0.0, 0.0, 0.0, 0.0, 0.0, 0.0,
	res: 4.216;	values: 1.0, 1.0, 1.0, 0.2, 0.0, 0.0, 0.0, 0.0, 0.0, 0.0,
	res: 4.221;	values: 1.0, 1.0, 1.0, 0.3, 0.0, 0.0, 0.0, 0.0, 0.0, 0.0,
	res: 4.224;	values: 1.0, 1.0, 1.0, 0.4, 0.0, 0.0, 0.0, 0.0, 0.0, 0.0,
	res: 4.225;	values: 1.0, 1.0, 1.0, 0.5, 0.0, 0.0, 0.0, 0.0, 0.0, 0.0,
	res: 4.224;	values: 1.0, 1.0, 1.0, 0.6, 0.0, 0.0, 0.0, 0.0, 0.0, 0.0,
	res: 4.221;	values: 1.0, 1.0, 1.0, 0.7, 0.0, 0.0, 0.0, 0.0, 0.0, 0.0,
	res: 4.216;	values: 1.0, 1.0, 1.0, 0.8, 0.0, 0.0, 0.0, 0.0, 0.0, 0.0,
	res: 4.209;	values: 1.0, 1.0, 1.0, 0.9, 0.0, 0.0, 0.0, 0.0, 0.0, 0.0,
	res: 4.200;	values: 1.0, 1.0, 1.0, 1.0, 0.0, 0.0, 0.0, 0.0, 0.0, 0.0,
Changing 5-th proportion
	res: 4.200;	values: 1.0, 1.0, 1.0, 1.0, 0.0, 0.0, 0.0, 0.0, 0.0, 0.0,
	res: 4.189;	values: 1.0, 1.0, 1.0, 1.0, 0.1, 0.0, 0.0, 0.0, 0.0, 0.0,
	res: 4.176;	values: 1.0, 1.0, 1.0, 1.0, 0.2, 0.0, 0.0, 0.0, 0.0, 0.0,
	res: 4.161;	values: 1.0, 1.0, 1.0, 1.0, 0.3, 0.0, 0.0, 0.0, 0.0, 0.0,
	res: 4.144;	values: 1.0, 1.0, 1.0, 1.0, 0.4, 0.0, 0.0, 0.0, 0.0, 0.0,
	res: 4.125;	values: 1.0, 1.0, 1.0, 1.0, 0.5, 0.0, 0.0, 0.0, 0.0, 0.0,
	res: 4.104;	values: 1.0, 1.0, 1.0, 1.0, 0.6, 0.0, 0.0, 0.0, 0.0, 0.0,
	res: 4.081;	values: 1.0, 1.0, 1.0, 1.0, 0.7, 0.0, 0.0, 0.0, 0.0, 0.0,
	res: 4.056;	values: 1.0, 1.0, 1.0, 1.0, 0.8, 0.0, 0.0, 0.0, 0.0, 0.0,
	res: 4.029;	values: 1.0, 1.0, 1.0, 1.0, 0.9, 0.0, 0.0, 0.0, 0.0, 0.0,
	res: 4.000;	values: 1.0, 1.0, 1.0, 1.0, 1.0, 0.0, 0.0, 0.0, 0.0, 0.0,
Changing 6-th proportion
	res: 4.000;	values: 1.0, 1.0, 1.0, 1.0, 1.0, 0.0, 0.0, 0.0, 0.0, 0.0,
	res: 3.969;	values: 1.0, 1.0, 1.0, 1.0, 1.0, 0.1, 0.0, 0.0, 0.0, 0.0,
	res: 3.936;	values: 1.0, 1.0, 1.0, 1.0, 1.0, 0.2, 0.0, 0.0, 0.0, 0.0,
	res: 3.901;	values: 1.0, 1.0, 1.0, 1.0, 1.0, 0.3, 0.0, 0.0, 0.0, 0.0,
	res: 3.864;	values: 1.0, 1.0, 1.0, 1.0, 1.0, 0.4, 0.0, 0.0, 0.0, 0.0,
	res: 3.825;	values: 1.0, 1.0, 1.0, 1.0, 1.0, 0.5, 0.0, 0.0, 0.0, 0.0,
	res: 3.784;	values: 1.0, 1.0, 1.0, 1.0, 1.0, 0.6, 0.0, 0.0, 0.0, 0.0,
	res: 3.741;	values: 1.0, 1.0, 1.0, 1.0, 1.0, 0.7, 0.0, 0.0, 0.0, 0.0,
	res: 3.696;	values: 1.0, 1.0, 1.0, 1.0, 1.0, 0.8, 0.0, 0.0, 0.0, 0.0,
	res: 3.649;	values: 1.0, 1.0, 1.0, 1.0, 1.0, 0.9, 0.0, 0.0, 0.0, 0.0,
	res: 3.600;	values: 1.0, 1.0, 1.0, 1.0, 1.0, 1.0, 0.0, 0.0, 0.0, 0.0,
Changing 7-th proportion
	res: 3.600;	values: 1.0, 1.0, 1.0, 1.0, 1.0, 1.0, 0.0, 0.0, 0.0, 0.0,
	res: 3.549;	values: 1.0, 1.0, 1.0, 1.0, 1.0, 1.0, 0.1, 0.0, 0.0, 0.0,
	res: 3.496;	values: 1.0, 1.0, 1.0, 1.0, 1.0, 1.0, 0.2, 0.0, 0.0, 0.0,
	res: 3.441;	values: 1.0, 1.0, 1.0, 1.0, 1.0, 1.0, 0.3, 0.0, 0.0, 0.0,
	res: 3.384;	values: 1.0, 1.0, 1.0, 1.0, 1.0, 1.0, 0.4, 0.0, 0.0, 0.0,
	res: 3.325;	values: 1.0, 1.0, 1.0, 1.0, 1.0, 1.0, 0.5, 0.0, 0.0, 0.0,
	res: 3.264;	values: 1.0, 1.0, 1.0, 1.0, 1.0, 1.0, 0.6, 0.0, 0.0, 0.0,
	res: 3.201;	values: 1.0, 1.0, 1.0, 1.0, 1.0, 1.0, 0.7, 0.0, 0.0, 0.0,
	res: 3.136;	values: 1.0, 1.0, 1.0, 1.0, 1.0, 1.0, 0.8, 0.0, 0.0, 0.0,
	res: 3.069;	values: 1.0, 1.0, 1.0, 1.0, 1.0, 1.0, 0.9, 0.0, 0.0, 0.0,
	res: 3.000;	values: 1.0, 1.0, 1.0, 1.0, 1.0, 1.0, 1.0, 0.0, 0.0, 0.0,
Changing 8-th proportion
	res: 3.000;	values: 1.0, 1.0, 1.0, 1.0, 1.0, 1.0, 1.0, 0.0, 0.0, 0.0,
	res: 2.929;	values: 1.0, 1.0, 1.0, 1.0, 1.0, 1.0, 1.0, 0.1, 0.0, 0.0,
	res: 2.856;	values: 1.0, 1.0, 1.0, 1.0, 1.0, 1.0, 1.0, 0.2, 0.0, 0.0,
	res: 2.781;	values: 1.0, 1.0, 1.0, 1.0, 1.0, 1.0, 1.0, 0.3, 0.0, 0.0,
	res: 2.704;	values: 1.0, 1.0, 1.0, 1.0, 1.0, 1.0, 1.0, 0.4, 0.0, 0.0,
	res: 2.625;	values: 1.0, 1.0, 1.0, 1.0, 1.0, 1.0, 1.0, 0.5, 0.0, 0.0,
	res: 2.544;	values: 1.0, 1.0, 1.0, 1.0, 1.0, 1.0, 1.0, 0.6, 0.0, 0.0,
	res: 2.461;	values: 1.0, 1.0, 1.0, 1.0, 1.0, 1.0, 1.0, 0.7, 0.0, 0.0,
	res: 2.376;	values: 1.0, 1.0, 1.0, 1.0, 1.0, 1.0, 1.0, 0.8, 0.0, 0.0,
	res: 2.289;	values: 1.0, 1.0, 1.0, 1.0, 1.0, 1.0, 1.0, 0.9, 0.0, 0.0,
	res: 2.200;	values: 1.0, 1.0, 1.0, 1.0, 1.0, 1.0, 1.0, 1.0, 0.0, 0.0,
Changing 9-th proportion
	res: 2.200;	values: 1.0, 1.0, 1.0, 1.0, 1.0, 1.0, 1.0, 1.0, 0.0, 0.0,
	res: 2.109;	values: 1.0, 1.0, 1.0, 1.0, 1.0, 1.0, 1.0, 1.0, 0.1, 0.0,
	res: 2.016;	values: 1.0, 1.0, 1.0, 1.0, 1.0, 1.0, 1.0, 1.0, 0.2, 0.0,
	res: 1.921;	values: 1.0, 1.0, 1.0, 1.0, 1.0, 1.0, 1.0, 1.0, 0.3, 0.0,
	res: 1.824;	values: 1.0, 1.0, 1.0, 1.0, 1.0, 1.0, 1.0, 1.0, 0.4, 0.0,
	res: 1.725;	values: 1.0, 1.0, 1.0, 1.0, 1.0, 1.0, 1.0, 1.0, 0.5, 0.0,
	res: 1.624;	values: 1.0, 1.0, 1.0, 1.0, 1.0, 1.0, 1.0, 1.0, 0.6, 0.0,
	res: 1.521;	values: 1.0, 1.0, 1.0, 1.0, 1.0, 1.0, 1.0, 1.0, 0.7, 0.0,
	res: 1.416;	values: 1.0, 1.0, 1.0, 1.0, 1.0, 1.0, 1.0, 1.0, 0.8, 0.0,
	res: 1.309;	values: 1.0, 1.0, 1.0, 1.0, 1.0, 1.0, 1.0, 1.0, 0.9, 0.0,
	res: 1.200;	values: 1.0, 1.0, 1.0, 1.0, 1.0, 1.0, 1.0, 1.0, 1.0, 0.0,
Changing 10-th proportion
	res: 1.200;	values: 1.0, 1.0, 1.0, 1.0, 1.0, 1.0, 1.0, 1.0, 1.0, 0.0,
	res: 1.089;	values: 1.0, 1.0, 1.0, 1.0, 1.0, 1.0, 1.0, 1.0, 1.0, 0.1,
	res: 0.976;	values: 1.0, 1.0, 1.0, 1.0, 1.0, 1.0, 1.0, 1.0, 1.0, 0.2,
	res: 0.861;	values: 1.0, 1.0, 1.0, 1.0, 1.0, 1.0, 1.0, 1.0, 1.0, 0.3,
	res: 0.744;	values: 1.0, 1.0, 1.0, 1.0, 1.0, 1.0, 1.0, 1.0, 1.0, 0.4,
	res: 0.625;	values: 1.0, 1.0, 1.0, 1.0, 1.0, 1.0, 1.0, 1.0, 1.0, 0.5,
	res: 0.504;	values: 1.0, 1.0, 1.0, 1.0, 1.0, 1.0, 1.0, 1.0, 1.0, 0.6,
	res: 0.381;	values: 1.0, 1.0, 1.0, 1.0, 1.0, 1.0, 1.0, 1.0, 1.0, 0.7,
	res: 0.256;	values: 1.0, 1.0, 1.0, 1.0, 1.0, 1.0, 1.0, 1.0, 1.0, 0.8,
	res: 0.129;	values: 1.0, 1.0, 1.0, 1.0, 1.0, 1.0, 1.0, 1.0, 1.0, 0.9,
	res: 0.000;	values: 1.0, 1.0, 1.0, 1.0, 1.0, 1.0, 1.0, 1.0, 1.0, 1.0,
Result
	Best value: 4.225
	Achieved at p = 3.5
		which defines 4-th round and investment of 0.5
	Max payout: 16.9 USD
\end{minted}

\end{document}